\documentstyle[11pt]{article}
\title{\vspace{2cm}\Large\bf \begin{center}
The local Hubble flow:
\end{center}
\begin{center}
A manifestation of dark energy
\end{center}}
 \author{
 Yurij Baryshev(1), Arthur Chernin(2,3), and
 Pekka Teerikorpi(2) \\
 (1) Institute of Astronomy, St.Petersburg State University,\\
 Staryj Peterhoff,198504, St.Petersburg, Russia\\
 (2) Tuorla Observatory, \\
 University of Turku, FIN-21500 Piikki\"o, Finland\\
 (3) Sternberg Astronomical Institute,\\
 Moscow University, 119899, Moscow, Russia.}

\date{~}

\begin{document}

\maketitle

\begin{abstract}

Our local environment at $r<10$ Mpc expands linearly and smoothly, as if
ruled by a uniform matter distribution, while
observations show the very clumpy local universe.
This is a long standing enigma in cosmology.
We argue that the recently discovered vacuum or quintessence
(dark energy (DE) component with
the equation of state $p_Q = w \rho_Q c^2$, $w \in [-1,0)$) from observations
of the high-redshift universe may also manifest itself in the
properties of the very local Hubble flow.
We introduce the concept of the critical distance $r_Q$ where
the repulsive force of dark energy starts to dominate over the gravity of
a mass concentration.  For the Local Group $r_Q$ is about $1.5$ Mpc.
Intriguingly, at the same distance $1.5$ Mpc
the linear and very "cold" Hubble flow emerges, with about the global
Hubble constant.
We also consider the critical epoch $t_Q$,
when the DE antigravity
began to dominate over the local matter gravity for a galaxy which
at the present epoch is in the local
DE dominated region.
Our main result is that the homogeneous dark energy component,
revealed by SNIa observations,
resolves the old confrontation
between the local Hubble flow
and local highly non-uniform, fractal matter distribution.
It explains why the Hubble law starts
on the outskirts of the Local Group, with the same Hubble constant as
globally and with a remarkably  small velocity dispersion.

\end{abstract}

\section{Introduction}

The Hubble diagram for high redshift Type Ia
Supernovae (Riess et al. 1998, Perlmutter et al. 1999)
and the Boomerang and MAXIMA-1
measurements of
the first acoustic peak location in
the angular power spectrum of the cosmic microwave background (CMB)
(de Bernardis 2000; Jaffe et al. 2000)
strongly favor a standard
cosmological model having the critical density ($\Omega = \Omega_m +
\Omega_{\Lambda} = 1$) and a dominant $\Lambda$-like,
``dark energy' component at the
present epoch ($\Omega_{\Lambda}^0 \approx 0.7$).
In this Letter we argue that the cosmological $\Lambda$ component is also
important for understanding the very local Hubble diagram. Namely,
there is a recognized riddle
 in observational cosmology
-- the linear and quiet Hubble flow in the very clumpy local universe
(Sandage et al. 1972; Weinberg 1977; Peebles 1992; Baryshev 1994;
Karachentsev \& Makarov 1996; Baryshev et al. 1998; Sandage 1999).

We show that
the Hubble flow emerges there, at a distance of about
1.5 Mpc, where the antigravity of the dark energy
starts to dominate over the gravity of clumpy matter.
We suppose that this is not a coincidence but
the key to the riddle of the local Hubble law.

\section{The enigma of the local Hubble flow}

\subsection{Formulation of the problem}

In the standard cosmology the dynamics described by
the Hubble law, $V = H\times r$, is a strict consequence of
a uniform distribution of self-gravitating
matter (Robertson 1955; Peebles 1993).
Hence the predicted linear velocity field is only valid for scales
where the universe is uniform.
However, a very puzzling fact, long ago noted by Sandage, Tammann
\& Hardy (1972; STH), is that
Hubble discovered his law in the distance interval $1-20$ Mpc, where
the galaxies are very clumpy distributed. Indeed, this  was
deep inside a cell of uniformity, the size of which is still debated,
but there are modern estimates from 30 Mpc up to 200 Mpc (Sylos Labini,
Montuori \& Pietronero 1998; Wu, Lahav \& Rees 1999)
Inside
 such cells the galaxies are fractally distributed,
 being clumpy at all smaller scales.
This means that "in fact, we would not  expect
any neat relation of proportionality
between velocity and distance for these [closeby] \ldots galaxies" (
Weinberg 1977).

On the other hand, recent studies of the local volume
($< 10$ Mpc), based on accurate distances, confirm
a "cold" linear Hubble flow, with about
the global Hubble constant $H_0$ (Sandage 1986; Karachentsev \& Makarov
1996; Teerikorpi 1997; Giovanelli et al. 1999: Ekholm et al. 1999).
Sandage (1999) recently expressed this surprising situation, as
the "extremely local rate [of expansion] is the same as the global rate
to better than $10\%$",  and an "explanation of why the local
expansion field is so noiseless remains a mystery".
Indeed, the velocity dispersion around
the local Hubble law is very low ($\sigma_V \leq 50-70$
km/s) (Sandage 1986; Karachentsev \& Makarov 1996; Ekholm et al. 2000;
in the last work Cepheid distances gave $\approx 40$ km/s).
  That here is a real problem, has been also shown with N-body
simulations
(Governato et al. 1997): the expected
velocity scatter close to galactic clumps like our LG is
$150-700$ km/s, depending on the CDM model.

One may thus divide the enigma of the local Hubble flow into
the following Problems:

(1) The linear velocity law appears at distances (1.5 Mpc) which are
about 1 percent of the scale where the Universe may finally appear uniform.
Why does our highly non-uniform closeby environment expand as if it were
uniform?  Why is the rate of expansion the same locally and globally,
i.e. deep inside and well outside of the cell of uniformity?

(2)  What makes the local Hubble flow so quiet?  Not only does the
Hubble law exist, but it has a remarkably small scatter.

\subsection{Previous attempts to explain the Hubble enigma}

In their seminal paper, STH discussed the hierarchical
galaxy distribution suggested by de Vaucouleurs (1971).
Wertz (1971) and Haggerty \& Wertz (1971) had made calculations on how
the Hubble law deviates from linearity inside such a structure.
STH did not find the predicted non-linearity and rejected the hierarchical
model.  However, they emphasized that the co-existence of the linear
Hubble law and the large local clouds of inhomogeneous matter still
is a dilemma.
STH proposed two possible solutions.
The linearly expanding universe may be inhomogeneous if its mean density is
 very low, $<< \rho_{crit}$. Or it may be filled by
 a totally uniform dark matter.

Modern redshift surveys, together with accurate measurements of the local
velocity field, have sharpened
this "Hubble - de Vaucouleurs" paradox (Baryshev et al. 1998).
Indeed, the 3-dimensional galaxy maps at scales
from 0.1 Mpc up to
20-200 Mpc are well described by a scale invariant fractal distribution
(Davis \& Peebles 1983; Pietronero 1987; Peebles 1993;
 Sylos Labini, Montuori \& Pietronero 1998;
 Teerikorpi et al. 1998; Wu, Lahav \& Rees 1999).
\footnote{Note that Mandelbrot (1982) replaced
the old concept of hierarchical galaxy distribution by the
more adequate concept of stochastic fractals.}
Using new data on large scale
structure and on the Hubble law,
the solutions mentioned by STH
were studied in Baryshev et al. (1998).
>From the linear theory of the gravitational growth of density
fluctuations (without dark energy),
 they concluded that either $\Omega_m \leq 0.01$ or for the uniform
dark matter $\Omega_{dark} \geq 0.99$.


 Nevertheless, these explanations are not sufficient, if considered
together with recent cosmological observations
which give $\Omega_m \approx 0.3$.
Also there is no direct evidence for a high-density uniform dark
  matter. Hence the riddle still exists.

\section{Dark energy and dynamical model}

\subsection{Vacuum, quintessence, dark energy}

 Recent detection of the dark energy component provide a natural
 candidate for a high-density truly uniform background, giving a novel
 possibility for resolving the Hubble enigma.

  In the cosmological
theory there are several kinds of dark energy with
a positive energy density ($\epsilon_Q = \rho_Q c^2 > 0$)
and negative pressure ($p_Q < 0$).
The dark energy or quintessence is a common name for: the classical
Einstein's cosmological constant $\Lambda$, time-variable
cosmological constant $\Lambda(t)$, cosmological
vacuum, effective scalar field and other forms of exotic matter
with negative pressure (see reviews by Bahcall et al. 1999; Sahni \&
Starobinsky 2000).
Dark energy is now much discussed and
used for interpretation of observational
cosmological tests (Wang et al. 2000; Podariu \& Ratra 2000).
The main reason to consider evolving DE is
the ``cosmic coincidence'': Why is
 the rapidly decreasing energy density of
matter at the present epoch so close to that of the
unchanging vacuum density?

The most striking property of the dark energy having the
equation of state $p_Q = w\epsilon_Q$, with $w \in [-1,0)$,
is that its gravitating mass
\begin{equation}
 M_Q
 = \frac{4\pi}{3} (1+3w)\rho_Q r^3
\label{mde}
\end{equation}
 is negative for $w < -1/3$.
This produces cosmological "antigravity" or a repulsion which
 accelerates the expansion of the universe.
E.g. for Einstein's
$\Lambda = 8 \pi G \rho_{\Lambda}/c^2$ (cosmological vacuum
with $w=-1$), the gravitating mass is  $M_{\Lambda} =
-8\pi \rho_{\Lambda}r^3 /3$.

\subsection{Dynamical model and the critical distance $r_Q$ }

We consider the dark energy component uniformly spreading everywhere, and
existing alongside with luminous and dark matter in the immediate
environment of our Galaxy.
Hence, the local dynamics is determined by the
competition of the gravity of dark (and luminous) matter and the antigravity
 of the DE.
The mass-to-luminosity ratios in systems of different scales
suggest that $M/L$ remains
constant for $r > 0.5$ Mpc (Bahcall, Lubin, \& Dorman 1995), hence dark
 matter should be distributed like luminous matter on such scales.
 The correlation analysis of
the space distribution of galaxies shows that
the mean density of luminous matter (and hence, of dark matter) decreases
with increasing scale so that the matter mass increases as
$M_m (r) \propto r^D$, where $D$, the fractal dimension,
is between 1 and 2 at least on scales up to 20 Mpc
(Wu, Lahav \& Rees 1999 ).
As its density is constant ($M_Q(r) \propto
r^3$), the DE component starts to dominate after some distance $r_Q$.

For a simple estimate of the scale beyond which
vacuum or quintessence dominates, we consider two models that
may imitate the environment of
the Local Group, where the local Hubble law is accurately observed.
The first model is a generalization of Sandage's
(1986) point-mass model, where a mass $M$ is
placed on the dark energy background with density
$\rho_Q$.
In the second model, a spherical distribution of
mass $M_m (r)$ is placed on the DE background.

The dynamics of expansion is described by the equation of motion
which is the (1,1)-component
of Einstein's field equations under the assumptions
of spherical symmetry, dust-like matter, and cosmological dark energy
given
by the energy-momentum (EM) tensor $T_k^i = \rho_Q c^2\; diag(1,-w,-w,-w)$:
\begin{equation}
\ddot{r} = - G M_{eff}/r^2; \;\;\; M_{eff} = M_m (r) + M_Q(r)\;.
\label{accel}
\end{equation}
Here $M_m (r) = M$, or $M_m(r) = M_*(r/r_*)^D$, respectively
for the first and second model, $M_Q$ is given by Eq.(\ref{mde}).

For the point-mass the  critical distance $r_Q$
 corresponding
to $\ddot{r}=0$, i.e. the distance where
DE antigravity compensates matter gravity, is defined by:
\begin{equation}
r > r_Q = (3M/(4\pi \tilde{w} \rho_Q))^{1/3}
\label{rQ1}
\end{equation}
where $\tilde{w} \equiv -3w-1$.
For $w \geq -1/3$ there is no repulsive force and $r_Q$
does not exist.

For the second model the dark energy term dominates dynamically at distances
\begin{equation}
r > r_Q = (r_*)^{-\frac{D}{3-D}} (M_*)^{\frac{1}{3-D}}
(\frac{3}{4\pi \tilde{w} \rho_Q})^{\frac{1}{3-D}}
\label{rQ2}
\end{equation}

\subsection{The temporal behavior of $r_Q$ and critical time $t_Q$}

In the case of cosmological constant a galaxy
which now is on the border of the vacuum dominated
region around a mass clump, previously lied inside the gravity
dominated sphere.
In terms of comoving coordinates, $r_Q$ shifts outwards for
 increasing redshift.
This prompts us to ask how $r_Q$ behaves
if the DE density changes, as
in quintessence models, where $r_Q$
may still shift outwards in comoving coordinates, but slower than for
the constant vacuum.

We illustrate the shift of $r_Q$ in time for three types
of DE models:
1) Einstein's cosmological constant $\Lambda$, 2) the quintessence
with $w = -2/3$, and 3) the case of ``coherent''
evolution of a quintessence
($\rho_Q = k\rho_{hm} $, where $\rho_{hm} $ is the uniform
matter component) during the late history of the universe,
with $w = -2/3$, $k = 1$.
The condition that the covariant divergence
of the total EM tensor (quintessence + matter) is zero
implies that
$\dot{\rho} = -3(\rho + \frac{p}{c^2})\dot{a}/a $,
where $\rho = \rho_Q + \rho_m$ and $p = p_Q $
for dust-like matter ($p_m =0$).
Hence the DE density behaves as
\begin{equation}
\rho_Q \propto a^{-3(\frac{1+k+wk}{1+k})}
\label{rhoQ}
\end{equation}
For $1/k = 0$ (no homogeneous matter) Eq.(\ref{rhoQ})
gives the usual (non-coherent) quintessence behavior
$\rho_Q \propto a^{-3(1+w)}$.

Let us consider
the ratio between two metric distances $r_{gal}/r_Q$, where
$r_{gal}$ is the distance to a galaxy which now is in the DE
dominated region and takes part in the Hubble flow.  The distance
$r_{gal}$ is
simply proportional to $a$.  The behavior of the critical distance
 $r_Q$ depends on the DE and matter models.
E.g. for the point-mass model the ratio is
$r_{gal}/r_Q = a^{(-w/(1+1/k))}$, which for $1/k=0$ gives
$a^{-w}$,
corresponding either to Einstein's $\Lambda$ ($w=-1$) or
to the quintessence (for $w \leq -1/3$).

We define a characteristic time for
the antigravity dominance.
It is counted from
the critical epoch $t_Q$, corresponding to the critical scale factor
 $a_Q = a (t=t_Q)$ when DE antigravity starts to dominate
over the gravity of matter concentration
component for a galaxy which
presently ($t=t_0$) is at the distance $r= r_{gal}=2r_Q^0$, i.e. well
within the DE dominated region.
E.g. for Sandage's point-mass model
thus defined critical scale factor is
\begin{equation}
 a_Q = 2^{(1+1/k)/w}
\label{aQ}
\end{equation}
With $1/k =0$ this gives $a_Q = 0.5$ for $w=-1$ and $a_Q=0.354$ for $w=-2/3$,
while with $k=1$,$w=-2/3$ the critical epoch is at $a_Q = 0.125$.
This illustrates the difference in critical times for different quintessence
models.

\section{ Specific properties of dark energy dominated dynamics }

\subsection{Structure evolution in DE dominated regions}

The theory of structure formation in the general case of an evolving
DE component is still under construction, though there are some results
for particular cases (Sahni \& Starobinsky 2000; Fabris \& Goncalves 2000).
If during the structure formation period most of the energy of the universe
produces antigravity and hence resists gravitational collapse, it is
impossible for the structure to grow at all. On the other hand, if antigravity
dominates only a short period, as in the case of the cosmological constant,
 then even
for $\Omega_{\Lambda} \sim 1$ it has only a slight effect on the infall
velocities around growing structures (Peebles 1984; Lahav et al. 1991).
This is clear from Eq.(\ref{accel})
which gives for the ratio of DE and point-mass
accelerations $g_Q/g_m \propto a^{-3w}$, hence for $w = -1$ the dynamical
influence of the constant $\Lambda$-component decreases very fast, while
for $w = -2/3$ the antigravity acts longer.

Another important consequence of the DE component is that within
$\Lambda$ dominated regions initial peculiar velocities decrease
due to adiabatic cooling.
The linear analysis of gravitational instability, following
Zeldovich (1965), shows that
 only decreasing modes exist in the vacuum-dominated region (Chernin,
Teerikorpi \& Baryshev 2000). This property of the accelerated
universe is important for understanding the local Hubble flow.
Our simple dynamical models led to the notion
of the distance $r_Q$ to the border between the matter and
DE dominated
regions. If a galaxy now beyond $r = r_Q$ also
 in the past was long enough
in the  DE dominated region, this would readily
 explain the smooth Hubble law at such distances.
 Then initial peculiar
velocities induced by the mass
of the Local Group (and similar more distant groups) have decreased
because of the adiabatic cooling ($\delta V(now) = \delta V(z)a(z)$).

Therefore a compromise situation is possible when
the time-variable DE is dominant during a period,
sufficient  for the structure growth in early times
and for the cooling of velocity dispersion in late times.
This means that such DE dominated regions between mass
concentrations are "pacific oceans" where the Hubble law appears and
the global Hubble constant may be measured even locally.

\subsection{Restoration of the Hubble flow in volumes with bulk motion}

We referred to two properties of the cosmological vacuum, namely uniformity
and antigravity. Let us now consider an important
third property of the vacuum.
It is clear that the
medium with the equation of state $p = - \rho c^2$ described by
the cosmological constant has EM tensor which
is proportional to the metric tensor $T_i^k = \alpha g_i^k$.
Hence it has the remarkable mechanical property of
vacuum: rest and motion can not be discriminated relative to it.
In other words, any matter motion is co-moving to this
medium.

When the vacuum dominates over the self-gravity of matter, matter
masses like galaxies  move as test particles
in the antigravity of the vacuum. The vacuum accelerates
the motions tending to form a regular kinematic pattern with
a linear velocity field. The expansion rate in such a flow
depends only on the value of the vacuum density.

These considerations can be applied to a local vacuum-dominated area,
no matter how fast is the bulk motion of the area against the CMB.
They are valid as well for the large-scale matter distribution;
in both cases, the flow will be linear and the rate of expansion
will be the same on any scale. And the flow will also be stable
against velocity or density perturbations.

\section{The critical distance $r_Q$ for the Local Group}

Let us calculate the critical distance for the Local Group
in the case of the point-mass model and the cosmological vacuum
 ($w=-1$ and $\rho_Q = \rho_{\Lambda}$ is
 constant in time).

The most recent
estimate for the mass of the LG gives
$M_{LG} =2\times 10^{12} M_{\odot} $ (van den Bergh 1999).
The vacuum density from SNIa observations is
 $\rho_{\Lambda} \simeq 0.7 \rho_{crit} =
  4.7 \times 10^{-30}h_{60}^2$ g/cm$^3$.
Then the distance $r_{\Lambda}$, where the vacuum starts to dominate,
 is $\simeq 1.5$ Mpc in the point-mass model.
If one changes $M_{LG}$ or $\rho_{\Lambda}$ by a factor of two,
 the critical distance is
changed by 26 percent.
If one chooses the mass distribution model with fractal dimension $D=1$
and $M_* = 2 \times 10^{12} M_{\odot}$ at $r_* =1$ Mpc, then Eq. (4)
gives $r_{\Lambda} = 1.8$ Mpc.

Hence, $r_{\Lambda}$ is robustly put
into the range from 1 to 2 Mpc.
This value is surprisingly close to the distance where the Hubble
law emerges, which is 1.5 Mpc (Sandage 1986).
Is this just a coincidence?  Not at all, we suggest  that this is
 rather a
major feature of the local matter flow. Indeed,
the dynamical dominance of
the dark energy at 1-2 Mpc and beyond means that the homogeneous vacuum
provides the dynamical background
for the Hubble flow of matter, on these distances.

\section{The local and global values of the Hubble constant}

Now we can understand Problem I -- the near-equality of the local and global
Hubble constants -- if the undistorted Hubble law is the signature of
the dynamical dominance of the uniform DE component.  In regions,
which have for a sufficient time been DE-dominated,
the rate of expansion -- in the first (and main!) approximation -- is
expressed via the DE density:
$H_0  \approx (8 \pi G\rho_Q/3)^{1/2}$.
The lumpy, fractally distributed matter gives only a small correction
to this formula for the global Hubble constant, because its average
density is much less than the DE density.  The very interesting aspect
of the dominating uniform DE  component is that
it allows us to predict the value of the Hubble constant from a local
estimate of the dark energy density.

If we identify the distance $r_{start}$ where the Hubble law starts,
with the critical radius $r_Q$ for our Local Group, then the
predicted value of $H_0$ is
\begin{equation}
H(predicted) \approx (\frac{2GM_{LG}}{\tilde{w}})^{1/2} (r_{start})^{-3/2}
\label{HQ}
\end{equation}
where $M_{LG}$ is the mass of the Local Group.
With the "standard" values $M = 2\times 10^{12} M_\odot$ and $r_{start} =
1.5$ Mpc, the Eq.(\ref{HQ}) gives for $H_0$ values from 53 km/s/Mpc
($w=-1$) to
74 km/s/Mpc ($w=-2/3$), covering the values
obtained for $H_0$ by different research groups. More conservatively, these
are upper limits, if in fact $r_Q < r_{start}$.
  We regard
this agreement with direct measurements of $H_0$ as evidence
for the cosmological vacuum and quintessence models. Note that
this reasoning for the DE component is based on locally
measured quantities ($M_{LG}$ and $r_{start}$) and is independent of
the supernovae determination of $\Omega_{\Lambda}$
(which did not depend on the value of $H_0$).

\section{The quietness of the local Hubble flow}

The possibility to solve Problem II may be illustrated by the following
simplified reasoning. At any epoch $a(t)$ and on the scale $r$ one
may regard $\delta V \approx Hr$ as an upper limit to the scatter around
the Hubble law.
For models  considered in Sec.3.3  the curvature is zero and hence
from Eq.(\ref{rhoQ}) it follows that
the Hubble constant $H(a)$ depends on the scale factor as
$a^{-(3/2)\beta }$, where $\beta = (1+k+wk)/(1+k)$.
For example in the case of vacuum, i.e. $w=-1$ and $1/k=0$ we have
$H=const$, and in the case of dust matter $k=o$ hence
$H \propto a^{-3/2}$.

So the scatter $\delta V(a)
< H_0 r_0 ~a^{1-(3/2)\beta}$, where $r_0$ is the considered scale at the
present epoch. E.g. for $r_0 = 5$ Mpc and $H_0 = 60$ km/s/Mpc,
$\delta V(a) < 300 ~a^{1-(3/2)\beta}$ km/s. This is also valid at the
critical epoch $a_Q$ after which the peculiar velocities within
a DE dominated region start to cool down adiabatically,
so that $\delta V(now) = \delta V(a_Q)\; a_Q$.
Hence one expects now
$\delta V(now) < 300 ~a_Q^{2-(3/2)\beta}$ km/s.

For simple numerical examples we use
the critical epoch $a_Q$ calculated in Sec.3.3.
In the case of the cosmological constant we have
$a_Q = 0.500$, hence $\delta V(now) < 75$ km/s.
  For the quintessence with $w = -2/3$,
 $a_Q = 0.354$ and $\delta V(now) < 63$ km/s, and
for the case of coherent evolution
$w = -2/3$, $k=1$,
the critical scale factor $a_Q = 0.125$ and now one
expects to observe $\delta V(now) < 38$ km/s.

Of course, these numbers  just illuminate the trend
for different DE models. A complete
treatment of the dispersion problem needs high-resolution large-volume
simulations of structure formation on the evolving DE background.

\section{Conclusions}

We conclude that a cosmological dark energy component
allows one to understand the emergence of the linear and quiet Hubble flow at
the border of the Local Group. The coincidence of the $r_Q$ with
the starting distance of the Hubble law
is seen to have the same roots as the ``cosmic coincidence''
of $\Lambda$ and matter densities. A summary of our results is:

* We introduce the concept of the critical distance
$r_Q$ where the
repulsive force starts to dominate over the gravity of
matter.  For the Local Group $r_Q \approx 1.5$ Mpc
in the case of $\rho_Q = 0.7 \rho_{crit}$.  This is
close to the distance where
the local Hubble flow emerges.
We make this new cosmic coincidence as a starting
point for understanding the cold local Hubble flow as reflecting
the dark energy dominated region around us.

* In dark energy dominated regions of the universe ("pacific oceans")
the linear Hubble law exists
due to the homogeneity of the DE component, with the same Hubble
 constant at first approximation
determined by the dark energy density.  This resolves Problem I
that the Hubble law not only starts at the border of the
DE dominated
region, but it appears with the $H_0$ as expected from the
dark energy density.

* The solution of Problem II
follows from the fact that a galaxy spending a sufficient time
in the DE dominated region
looses its initial peculiar velocity which cools down adiabatically.
The mysterious quietness of the local Hubble flow does not
look at all dramatic in the context of the dark energy driven expansion.

* Our results support a dark energy component which
presently dominates
the dynamics of the Universe on scales from the Local Group to the
Hubble radius. Even the closeby galaxy universe is now seen as a cosmic
laboratory where all the physical ingredients of the universe:
luminous matter, dark matter, and the dark energy, may be
detected.
Hence, the very local volume is extremely important for the study of global
 properties of the universe.
 Cosmology starts immediately beyond the border of
the Local Group!

In other environments the DE dominance may start from
smaller or larger distances. So, around the Coma
cluster, the vacuum starts dominating from about 20 Mpc.
For a cosmologist in Coma, cosmology begins around such a distance.
In such DE dominated regions the quiet Hubble flow
is produced by the same vacuum density and hence the Hubble constant is the
same.

 One now realizes that when Hubble found his linear redshift law in
the closeby lumpy environment,
he actually saw the influence of the mysterious uniform dark energy.
Though
the value of the vacuum or DE density is still poorly understood
in cosmological theory,
the observations make one agree with the old words by Lema\^{i}tre
who was the first to associate $\Lambda$
with the vacuum having positive energy density
and negative pressure leading to a cosmological repulsion:
"Everything happens as though the energy {\em in vacuo} were different
from zero." (Seitter \& Duemmler 1989).

{\Large \bf Acknowledgements}

We are grateful to Chris Flynn for helpful comments.
This study has been supported by The Academy of Finland (project
``Cosmology in the local galaxy universe'').

\newpage

{\LARGE \bf \begin{center}
References
\end{center}}

Bahcall, N.A., Lubin, L.M., Dorman, V. 1995, Astrophys.J., {\bf 447}, L81

Bahcall N., Ostriker J., Perlmutter S., Steinhardt P. 1999,
Science {\bf 284}, 1481

Baryshev Yu. 1994, Astron.Astrophys.Trans.,{\bf 5}, 15

Baryshev Yu., Sylos Labini F., Montuori M., Pietronero L., Teerikorpi P. 1998,
Fractals, {\bf 6}, 231


Chernin A., Teerikorpi P., Baryshev Yu. 2000, Adv. Space Res.(in press)


Davis M., Peebles P.J.E. 1983, ApJ, {\bf 267}, 465

de Bernardis et al. 2000, Nature, {\bf 404}, 955

de Vaucouleurs G. 1971, Science, {\bf 167}, 1203

Ekholm T., Lanoix P., Teerikorpi P., Paturel G., Fouqu\'{e} P. 1999,
A\&A, {\bf 351}, 827

Ekholm T., Baryshev Yu., Teerikorpi P., Hanski M., Paturel G. 2000,
A\&A (submitted)

Fabris, J.C., Goncalves, S.V.B., 2000, gr-gc/0010046

Giovanelli R., Dale D., Haynes M., Hardy E., Campusano L. 1999,
 ApJ, {\bf 525}, 25

Governato F. et al. 1997, New Astr., {\bf 2}, 91


Haggerty M., Wertz R.J. 1971, MNRAS, {\bf 155}, 495

Jaffe A.H. et al. 2000, astro-ph/0007333

Karachentsev, I., Makarov, D. 1996, AJ, {\bf 111}, 794

Lahav O., Lilje P., Primack J., Rees M. 1991, MNRAS, {\bf 251}, 126

Mandelbrot B. 1982 {\em The Fractal Geometry of Nature}
(W.H. Freeman )

Peebles P.J.E. 1984, ApJ, {\bf 284}, 439

Peebles P.J.E., 1992, Lecture in the symposium "The Cosmic
Microwave Background Radiation and the Large-Scale Structure of
the Universe after COBE" (Stockholm 1992)

Peebles, P.J.E. 1993 {\em Principles of Cosmology} (Princeton: Princeton
Univ. Press)


Perlmuter, S. et al. 1999, ApJ, {\bf 517}, 565

Pietronero L. 1987, Physica, {\bf A144}, 257

Podariu S., Ratra B. 2000, ApJ, {\bf 532}, L109

Riess, A.G. et al. 1998, AJ, {\bf 116}, 1009

Robertson, H.P. 1955, PASP, {\bf 67}, 82

Sahni V., Starobinsky A., 2000, astro-ph/9904398

Sandage, A. 1986, ApJ, {\bf 307}, 1

Sandage, A. 1999, ApJ, {\bf 527}, 479

Sandage A., Tammann G., Hardy E. 1972, ApJ, {\bf 172}, 253

Seitter, W.C., Duemmler, R. 1989, in {\em Morphological Cosmology}
(eds. P. Flin, H.W. Dverbeck), Springer, Berlin, pp. 377-387

Sylos Labini, F., Montuori, M., Pietronero, L. 1998,
Phys.Rep., {\bf 293}, 61

Teerikorpi, P. 1997, ARA\&A, {\bf 35}, 101

Teerikorpi, P. et al. 1998, A\&A, {\bf 334}, 395

van den Bergh, S. 1999, A\&A Rev., {\bf 9}, 273

Wang L., Caldwell R.R., Ostriker J.P., Steinhardt P.J. 2000,
ApJ, {\bf 530}, 17

Weinberg, S. 1977 {\em The First Three Minutes} (Basic Books, New York), p.26

Wertz J.R. 1971, ApJ, {\bf 164}, 227

Wu, K.K.S., Lahav, O., Rees, M.J. 1999, Nature {\bf 397}, 225

Zeldovich, Ya.B. 1965, Advan. Astron. Ap., {\bf 3}, 241

\end{document}